%% file: main.tex
\documentclass[sigconf,natbib=true,anonymous=false]{acmart}
 
\usepackage{enumitem}
\usepackage{algorithm}
\usepackage{algpseudocode}
\usepackage{todonotes}
\usepackage{float}
\newfloat{algorithm}{t}{lop}

\usepackage{marginnote}
\newcommand{\pageenlarge}[1]{\enlargethispage{#1\baselineskip}}

\AtBeginDocument{%
  \providecommand\BibTeX{{%
    \normalfont B\kern-0.5em{\scshape i\kern-0.25em b}\kern-0.8em\TeX}}}

\copyrightyear{2023}
\acmYear{2023}
\setcopyright{acmlicensed}\acmConference[SIGIR '23]{Proceedings of the 46th International ACM SIGIR Conference on Research and Development in Information Retrieval}{July 23--27, 2023}{Taipei, Taiwan}
\acmBooktitle{Proceedings of the 46th International ACM SIGIR Conference on Research and Development in Information Retrieval (SIGIR '23), July 23--27, 2023, Taipei, Taiwan}
\acmPrice{15.00}
\acmDOI{10.1145/3539618.3591715}
\acmISBN{978-1-4503-9408-6/23/07}

\begin{document}

\title{Lexically-Accelerated Dense Retrieval}

\author{Hrishikesh Kulkarni}
\affiliation{%
  \institution{Georgetown University}
  \city{Washington, DC}
  \country{USA}}
\email{first@ir.cs.georgetown.edu}

\author{Sean MacAvaney}
\affiliation{%
  \institution{University of Glasgow}
  \city{Glasgow}
  \country{UK}
}
\email{first.last@glasgow.ac.uk}

\author{Nazli Goharian}
\affiliation{%
 \institution{Georgetown University}
 \city{Washington, DC}
 \country{USA}}
\email{first@ir.cs.georgetown.edu}

\author{Ophir Frieder}
\affiliation{%
 \institution{Georgetown University}
 \city{Washington, DC}
 \country{USA}}
\email{first@ir.cs.georgetown.edu}

\renewcommand{\shortauthors}{Hrishikesh Kulkarni, Sean MacAvaney, Nazli Goharian, \& Ophir Frieder}
\begin{abstract}
Retrieval approaches that score documents based on learned dense vectors (i.e., dense retrieval) rather than lexical signals (i.e., conventional retrieval) are increasingly popular. Their ability to identify related documents that do not necessarily contain the same terms as those appearing in the user's query (thereby improving recall) is one of their key advantages. However, to actually achieve these gains, dense retrieval approaches typically require an exhaustive search over the document collection, making them considerably more expensive at query-time than conventional lexical approaches. Several techniques aim to reduce this computational overhead by \textit{approximating} the results of a full dense retriever. Although these approaches reasonably approximate the top results, they suffer in terms of recall -- one of the key advantages of dense retrieval. We introduce `LADR' (Lexically-Accelerated Dense Retrieval), a simple-yet-effective approach that improves the efficiency of existing dense retrieval models without compromising on retrieval effectiveness. LADR uses lexical retrieval techniques to seed a dense retrieval exploration that uses a document proximity graph. Through extensive experiments, we find that LADR establishes a new dense retrieval effectiveness-efficiency Pareto frontier among approximate $k$ nearest neighbor techniques. When tuned to take around 8ms per query in retrieval latency on our hardware, LADR consistently achieves both precision and recall that are on par with an exhaustive search on standard benchmarks. Importantly, LADR accomplishes this using only a single CPU -- no hardware accelerators such as GPUs -- which reduces the deployment cost of dense retrieval systems.
\end{abstract}



\maketitle

\begin{figure}[b!]
    \centering
    \includegraphics[width=0.495\linewidth]{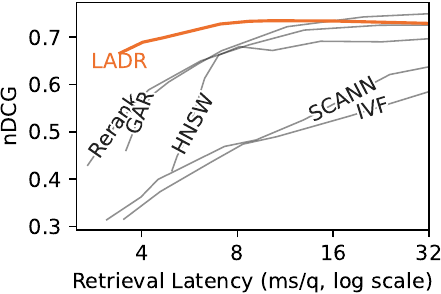}
    \includegraphics[width=0.495\linewidth]{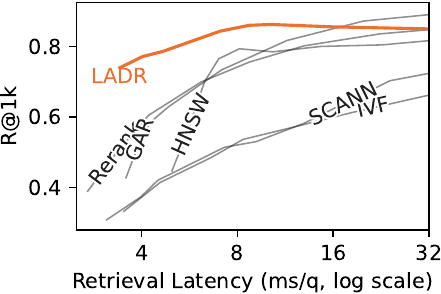}
    \vspace{-2em}
    \caption{Comparison of our approach (LADR) and baselines when retrieving using TAS-B~\cite{10.1145/3404835.3462891} on the TREC Deep Learning 2019 dataset, in terms of nDCG, Recall@1000, and latency (single CPU). LADR establishes a new Pareto frontier with higher nDCG and Recall at latencies below 16ms/query.
    }
    \label{fig:teaser}
\end{figure}

\section{Introduction}
To identify relevant documents, search engines typically make use of lexical and/or dense search operations. Lexical approaches use the terms that appear in a user's query to identify and score documents that contain the same terms (e.g., using an approach like BM25 over an inverted index). Lexical approaches like BM25 \cite{INR-019} fail to understand semantic relationships in different scenarios \cite{kim-etal-2021-query}. A major drawback of this approach is that it is limited to documents that use the exact terms that the user searches for, which can potentially affect the recall. Recall is particularly important when results are further re-ranked in a second stage since documents missed in the first stage cannot be identified in subsequent stages using the typical \textit{cascade} approach. To overcome this problem, \textit{dense retrieval} approaches --- which encode a user's query into a dense ``semantic'' vector and retrieve documents that were also encoded into semantic vectors --- are becoming increasingly popular. By relying on dense vector similarity measures, retrieved documents no longer need any term overlap with a user's query.

Such approaches come at a comparatively high query-time cost, however. To produce state-of-the-art results, an exhaustive search over all document vectors is usually required. This is significantly more expensive than lexical search since it scales linearly with the number of documents in the corpus. (Meanwhile, lexical search benefits from pruning the search space to only documents that use terms from the user's query, as well as other optimizations~\cite{10.1145/2009916.2010048}.) The efficiency problem of dense retrieval has spawned numerous efforts to \textit{approximate} top results.
Researchers proposed various approximation methods like HNSW \cite{8594636}, IVF \cite{1238663,DBLP:conf/eccv/YuanGCLJ12}, and ScaNN~\cite{avq_2020}. However, these methods have their own limitations. Most notably, they suffer in terms of recall, which is meant to be one of the key benefits of dense retrieval.

Remarkably, simply re-ranking an initial pool of lexical results (such as BM25) remains highly competitive~\cite{10.1145/3485447.3511955,10.1145/3477495.3531721}. We posit that this is due to two reasons. First, lexical signals are inherently valuable in many textual search tasks; intuitively, documents that satisfy a user's query will often contain the terms in the query itself. Second, pools of candidates with matching terms can be obtained very efficiently; decades of optimizations in this area have given us highly-optimized algorithms for the task, like BlockMax WAND~\cite{10.1145/2009916.2010048}. Nevertheless, using a lexical model as a pool of documents for re-ranking has inherent drawbacks, including exclusive reliance on lexical signals for retrieval --- another key problem that dense retrieval is meant to solve.

\pageenlarge{1}
In this paper, we propose a simple-yet-effective technique called Lexically-Accelerated Dense Retrieval (LADR). It builds on the strengths of existing approximate k-nearest neighbor approximation techniques by leveraging a pool of candidate documents retrieved using a lexical model to seed the dense retrieval search. By using the lexical document pool, LADR leverages highly-efficient lexical search to obtain reasonable seed documents. LADR then explores the neighboring documents in the dense retrieval model's semantic space, allowing for documents to be retrieved that do not contain lexical matches with the query. We explore two strategies for exploration: a \textit{proactive} strategy that greedily scores the neighbors of all seed documents, and an \textit{adaptive} strategy that iteratively scores neighbors of documents in the top results until they converge.

Through extensive experiments across several dense retrieval models on standard text retrieval test collections, we find that LADR establishes a Pareto frontier in terms of retrieval effectiveness and efficiency. For example, Figure~\ref{fig:teaser} compares the performance of LADR on the TAS-B~\cite{10.1145/3404835.3462891} model with alternative approximation techniques.\footnote{Each technique is parameterized by the degree of exploration. LADR and re-rank: number of candidates; HNSW: expansion factor; IVF/ScaNN: number of probes.} At all latencies between 4 and 16ms/query, we see that LADR exceeds the effectiveness of existing techniques in terms of nDCG and Recall@1000. Further analysis shows that the parameters that LADR introduces (e.g., the number of nearest neighbors in the proximity graph) allow the user to easily trade off efficiency, effectiveness, and storage requirements, and that the approach is largely robust to alternative sources of proximity signals, such as approximate proximity graphs or graphs based on lexical proximity.

In summary, our contributions are as follows:

\begin{itemize}[leftmargin=*]
\item We introduce LADR, a technique for reducing the computational overhead of dense retrieval while maintaining high effectiveness.
\item  We explore a proactive and an adaptive strategy to ensure effective exploration and optimal use of time budget for better retrieval.
\item We conduct extensive experiments on standard benchmarks and a variety of single-representation dense retrieval models and find that our approach establishes a new Pareto frontier for low-latency approximate dense retrieval.
\item Further analysis demonstrates that LADR is robust to its introduced parameters and alternative sources of document proximity.
\end{itemize}

\section{Related Work}\label{s:rw}
A significant number of AI problems can be modelled as search or information retrieval problems. Increasing quantum of information and need of very high efficiency always make this problem exceedingly challenging. Multiple approaches have been tried out to improve information retrieval effectiveness \cite{10.1561/1500000061}. Generally, frequency of word occurrence in a document gives important measurement of word significance \cite{5392672}. Typically, heuristic-based search approaches get the exploration directions from the knowledge about search space and query decoding.
Tree-based \cite{pmlr-v119-zhuo20a} and Product Quantization-based \cite{5432202,10.1145/3404835.3462988} indexing methods have been proposed.
Popularly, three types of information retrieval models have inspired traditional approaches viz  Boolean models, vector space models and probabilistic models \cite{liu2007web}. Thus,  different methods and models such as indexing, ranked retrieval, similarity measures and query expansion proposed by different researchers created a platform to solve problem of getting pertinent information challenged by information expansion.

\pageenlarge{1}
\subsection{Lexical Information Retrieval}

Generally traditional approaches for retrieval exploit frequency and occurrence of query terms in the target document text \cite{Mikolov2013EfficientEO}. In this approach the thrust is on exact term matching \cite{10.1145/2983323.2983769} creating basis for traditional IR systems. Varied weighting and normalization formulations over the frequency and occurrence of query terms in target documents \cite{10.5555/1861751.1861756} led to a variety of TF-IDF models exemplified by BM25 \cite{INR-019}. These bag-of-words based approaches like BM25 created initial avenues to negotiate with efficiency aspects of retrieval problem. Multiple variations of BM25 were introduced to tackle different application specific challenges \cite{INR-019}. 
Patterns identified in the query and documents should be independent of exact word matching and should highlight the relevance in case of different articulations of the same concept. 
To counter the issue of tight coupling of vocabulary to IR systems \cite{mitra2018an}, different dictionary-based and n-gram-based techniques evolved to query likelihood \cite{Xue2009QuerySB} and query expansion for effective retrieval. These techniques have offered initial ground for IR and later machine learning techniques offered serious technical leap to enhance these techniques. 
First-stage retrieval models like SPLADE learn effective and efficient sparse representations where the sparsity can be controlled through regularization \cite{10.1145/3404835.3463098}.
Guided processing heuristic is also used to boost efficiency in sparse models without significant loss in effectiveness \cite{10.1145/3477495.3531774}. 
Traditional NLP centric and token driven hand crafted IR evolved to more complex supervised learning models.  

\subsection{Neural Information Retrieval}

Enhanced traditional information retrieval methods have too much vocabulary dependence. These enhancements fail to match the higher semantic levels due to underlined constrained semantic framework \cite{https://doi.org/10.48550/arxiv.2009.01938}. The focus on inspecting query and ignoring all whereabouts coming from the documents, limits the effectiveness of this approach. 
Neural models produce better results with the use of dense text representations. Typically, two primary types of neural models exist \cite{https://doi.org/10.48550/arxiv.1606.04648}. One is interaction based where interactions between words in queries are modelled. On the other hand, representation-based models learn a single vector representation of the query. 
The neural IR models can be used in conjunction with other models \cite{yates-etal-2021-pretrained}. Bag-of-words models and other lexical methods can also be enhanced with machine learning models \cite{Askari2021CombiningLA}. This may include use of machine learning based query expansion for sparse features coming through bag-of-words or learning based term weight tuning where BERT like pre-trained transformer models could be used \cite{Dai2020ContextAwareTW}. Models like TAS-B~\cite{10.1145/3404835.3462891} and TCT-ColBERT-HNP~\cite{lin-etal-2021-batch} produce state of the art results in dense retrieval. Considering the efficiency-effectiveness tradeoffs between sparse and dense models, hybrid models were proposed and choice of model was made based on time and resource budget \cite{10.1145/3459637.3482159}. Sparse auto encoders have also been used to increase efficiency in building parallel inverted index in Siamese-BERT based models \cite{lassance2021composite}.
As complimentary information is provided by sparse and dense methods, `hybrid' methods have also been explored. Researchers combined sparse and dense retrieval ranking on the basis of relevance score interpolation leading to promising results \cite{10.1007/978-3-030-72113-8_10,journals/corr/abs-2010-01195,lin-etal-2021-batch}.

\pageenlarge{1}
\subsection{Re-ranking Pipeline to Improve Efficiency}

Traditional token-based and vocabulary-centric approaches like BM25 are efficient but not effective. On the contrary, dense retrieval approaches mainly based on contextualized transformer language models like BERT are highly effective but not efficient.
Many researchers worked on approaches to strike the balance between efficiency and effectiveness. The most popular technique is to perform a two stage retrieval \cite{https://doi.org/10.48550/arxiv.2111.13853}. Efficient lexical methods are used at the first stage to retrieve an initial pool of documents based on the re-ranking budget. Effective dense retrieval methods are used at the second stage to re-rank the obtained results at the first stage~\cite{10.1145/3477495.3531721,10.1145/3485447.3511955,macavaney:sigir2020-epic}. Re-ranking pipelines perform very well 
but are limited by recall. Our LADR approach builds off this idea by seeding dense exploration with the results from a first-stage lexical ranker. This is similar to an idea proposed by~\cite{macavaney:cikm2022-adaptive} for the purposes of finding additional documents to score when re-ranking; ours differs from this work because we are exploring similar documents as a method of pruning the search space of dense retrieval while keeping comparable performance to an exhaustive search.

\subsection{Seed-based Exploration in Proximity Graphs}

Hierarchical Navigable Small World graphs (HNSW) \cite{8594636} are used to perform efficient dense retrieval. Graph-based approaches use the clustering hypothesis which states that relevance among documents could be mapped to the same query \cite{Jardine1971TheUO}. Proximity graph methods exemplified by HNSW are current state of the art methods in ANN search and large scale retrieval using dense vector representations can be modelled as nearest neighbor search. To speed up the proceeding, approximate nearest neighbor search using HNSW is used. This results in substantial decrease in latency when compared to exhaustive dense retrieval. Another competitive index is Inverted Files index (IVF) \cite{1238663,DBLP:conf/eccv/YuanGCLJ12} which is based on Voronoi diagrams (Dirichlet tessellation). IVF also performs clustering to reduce the search scope. IVF based dense retrieval also reduces latency to a great extent delivering decent search-quality. ScaNN \cite{avq_2020} is another quantization based technique which works on anisotropic quantization loss function for inner product search.

\begin{figure}
\centering
\includegraphics[width=\linewidth]{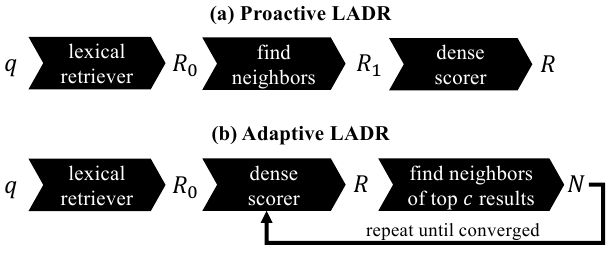}
\caption{Overview of (a) Proactive and (b) Adaptive LADR.}
\label{fig:arch}
\end{figure}

Most of the recent approaches couple random seed point-based exploration with dense retrieval. The initiation points are influenced by the proximity graph building algorithms and do not take differentiating query characteristics into account \cite{1238663,DBLP:conf/eccv/YuanGCLJ12}. Random exploration of search space until global minima is reached is often very expensive and the above mentioned approaches usually terminate around local minimas. Identification of right seed points to have informed exploration definitely could add value in terms of efficiency as well as accuracy. Multiple applications have witnessed enhancements in various aspects by using seeding techniques \cite{LECHTENBERG2022116967,8594636,10.1145/3341981.3344250}. These seed points could be better derived from input document or query. We empower dense retrieval models with bag-of-words based seeding approach for informed exploration. 

\pageenlarge{1}
\section{LADR}

Lexically Accelerated Dense Retrieval (LADR) is a simple technique that combines the strengths of two leading dense retrieval techniques: \textit{re-ranking}, which can efficiently identify potentially-relevant documents using lexical signals; and \textit{HNSW}, which limits the search space based on a pre-computed document proximity graph. LADR leverages an efficient lexical model to identify good ``seed'' documents. It then makes use of a document proximity graph to expand the search space to similar documents. Here, the document proximity graph is computed offline. For each document, neighbors, i.e., 128 closest documents, are determined using the dense vector document similarity method TCT-ColBERT-HNP \cite{macavaney:cikm2022-adaptive}. We identify two variants of LADR: a \textit{proactive} variant (Section~\ref{sec:method-pro}), which indiscriminately explores the neighbors of the seed documents, and an \textit{adaptive} variant (Section~\ref{sec:method-ada}), which selectively and iteratively explores the neighbors of the top $c$ documents until those results converge. Figure~\ref{fig:arch} provides an overview of the two approaches. Each variant has its own advantages, which we detail in the following sections.

\subsection{$R_0$ Lexical Seed Documents}

Both the proactive and adaptive variants make use of an initial seed document set $R_0$ obtained from the top $n$ results of a lexical retrieval model, e.g., BM25~\cite{INR-019}. Such models are largely seen as reasonable baselines for text retrieval tasks~\cite{10.1145/3308774.3308781}, and are generally robust across tasks and domains~\cite{thakur2021beir}. Beyond providing a competitive starting point in terms of recall, there are several key efficiency advantages in using a lexical retrieval method for generating seeds. First, decades of optimizations have resulted in highly-optimized query processing algorithms, such as BlockMax-WAND~\cite{10.1145/2009916.2010048}. These algorithms generally result in sub-linear query processing time with respect to both the number of documents indexed\footnote{For instance, PISA~\cite{Mallia2019PISAPI} retrieves the top 1000 BM25 results over 100k, 1M, and 10M passages in 0.4, 1.1, and 4.2 ms on average, respectively.} and the number of top results returned.\footnote{For instance, PISA~\cite{Mallia2019PISAPI} retrieves the top 10, 100, and 1000 BM25 results for 10M passages in 1.1, 1.9, and 4.2 ms on average, respectively.} 
They also are inexpensive to construct, compress, and store without specialized hardware acceleration, especially when compared to the overheads of dense retrieval models.\footnote{For instance, it takes PISA~\cite{Mallia2019PISAPI} around 2 minutes to build an optimized BM25 index for MS MARCO (727MB in size), while it takes around 3 hours to compute TAS-B~\cite{10.1145/3404835.3462891} vectors for MS MARCO using an A6000 GPU (26GB in size).}
Finally, it is commonplace in industry to use a lexical search index for first stage retrieval already (e.g., tools like ElasticSearch are prevalent), so the choice fits with existing operational structures.

\subsection{Proactive LADR}\label{sec:method-pro}

Now that seed documents have been selected, we explore strategies for identifying additional documents to score.
We first explore a \textit{Proactive} LADR technique (see Algorithm~\ref{alg:proladr}). This variant takes a na\"ive approach for choosing documents to score; it simply expands the initial seed documents into $R_1$, by taking the union of the seed document set ($R_0$) and the $k$ nearest neighbors of those documents. The final results are then selected by scoring $R_1$ using the dense scoring function (i.e., computing the similarity between the encoded query vector and the pre-computed document vectors associated with documents in $R_1$).

\pageenlarge{1}
The degree of approximation of Proactive LADR is controlled by $n$ (seed set size) and $k$ (number of neighbors). As $n$ increases, more potentially-relevant documents (or neighbors of potentially relevant documents) can be introduced by the lexical retriever. Meanwhile, as $k$ increases, every seed document casts a wider net, allowing for more distant (but still similar) documents to be scored. We explore the interplay of these parameters in Section~\ref{rq34}.

\textbf{Computational and Storage Overheads.} The computational cost of Proactive LADR is trivially bounded by the cost of lexical retrieval (usually sub-linear) plus $O(kn)$ vector similarity calculations. In practice, there is likely to be a substantial overlap between the neighbors of the seed documents due to the clustering hypothesis~\cite{Jardine1971TheUO}, further reducing the cost. Meanwhile, the storage overheads are the $n$ nearest neighbors of each document, $O(Dn)$, where $D$ represents the number of documents in the corpus.

\subsection{Adaptive LADR}\label{sec:method-ada}

Proactive LADR's na\"ive approach of selecting the neighbors of all seed documents potentially wastes compute by scoring the neighbors of documents that turn out to be not so relevant. On the other hand, Adaptive LADR (see Algorithm~\ref{alg:adaladr}) attempts to correct for this by adaptively choosing which neighbors to score based on the $c$ top-scoring documents seen up to that point. Adaptive LADR starts by scoring the initial seed set of documents $R_0$. It then finds and scores the $k$ nearest neighbors of the top $c$ results. The algorithm then repeats the process until no new documents are introduced into the top $c$ results. By being more selective about picking neighbors, Adaptive LADR can potentially spend the limited computational budgets (which are dominated by similarity scoring) on the most promising documents. When compared with the proactive strategy, it also makes it possible to move multiple hops in the document proximity graph, when a neighbor scores within the top $c$ results.

\textbf{Computational and Storage Overheads.} Unlike Proactive LADR, the worst-case computational cost of Adaptive LADR is the cost of lexical retrieval plus $O(D)$ vector similarity calculations, which assumes a very poor selection of seed documents, necessitating a full exploration of the document graph. This case is very unlikely, given the well-established reasonable effectiveness of lexical search models like BM25. Even so, practitioners may consider including a timeout or other suitable termination criteria in Adaptive LADR to conclude the search early; we leave the exploration of the effect of such strategies for future work. The storage overheads are the same as for Proactive LADR: $O(Dn)$ needed to store the document graph where n represents the number of neighbors stored for each document.

\begin{algorithm}
\caption{Proactive LADR}\label{alg:proladr}
\begin{algorithmic}
\Require $q$ query, $n$ seed set size, $k$ number of neighbors
\Ensure $R$ dense retrieval results
\State $R_0 \gets \Call{LexicalRetrieve}{q, n}$ \Comment{seed with lexical results}
\State $R_1 \gets R_0 \cup \Call{Neighbors}{R_0,k}$ \Comment{add neighbors}
\State $R \gets \Call{Score}{q, R_1}$ \Comment{score seeds and their neighbors}
\end{algorithmic}
\end{algorithm}

\begin{algorithm}
\caption{Adaptive LADR}\label{alg:adaladr}
\begin{algorithmic}
\Require $q$ query, $n$ seed set size, $c$ exploration depth, $k$ number of neighbors
\Ensure $R$ Dense Retrieval Results
\State $R_0 \gets \Call{LexicalRetrieve}{q,n}$ \Comment{seed with lexical results}
\State $R \gets \Call{Score}{q, R_0}$ \Comment{score seeds}
\State $N \gets \Call{Neighbors}{\text{top }c\text{ from }R, k}$ \Comment{neighbors of top results}
\State $N \gets N\setminus R $ \Comment{skip docs we've already scored}
\While{$|N| \neq 0$} \Comment{iterate until no new docs identified}
\State $R \gets R \cup \Call{Score}{N}$ \Comment{score and add neighbors}
\State $N \gets \Call{Neighbors}{\text{top }c\text{ from }R, k} \setminus R$
\EndWhile
\end{algorithmic}
\end{algorithm}

\pageenlarge{1}
\subsection{Comparison of Variants}

Proactive LADR expands the initial seed documents by taking into consideration the k nearest neighbors of each of the initial seed document. While Adaptive LADR keeps expanding the initial seed documents by considering k nearest neighbors selectively (only of top c documents) and iteratively (until convergence). 
Though relatively simple, Proactive LADR has some distinct advantages over more complicated LADR approaches. First, there is a trivial $O(kn)$ upper bound to the number of vector comparisons made which is $O(D)$ in the case of Adaptive LADR. In practice, the number of comparisons is going to be significantly less due to overlap in the neighbors of seed documents based on the clustering hypothesis and reasonable effectiveness of BM25. But in worst case scenario Adaptive LADR will be equivalent to exhaustive retrieval. 
As Proactive LADR is a pipeline approach, operations can be trivially delegated (e.g., a separate service might be responsible for computing vector similarities, which only needs to be invoked a single time for Proactive LADR). On the other hand Adaptive LADR operations tend to be slightly demanding and costlier due to the feedback loop introduced by the iterative exploration approach (e.g., a separate vector similarity service would need to be invoked multiple times, which could incur higher costs).

\pageenlarge{1}
\subsection{Comparison with Existing Methods}

\textbf{Re-ranking.} Both LADR variants build upon the high effectiveness of the re-ranking technique~\cite{10.1145/3485447.3511955,10.1145/3477495.3531721}. Using a lexical model for re-ranking alone is a major drawback, however: it inherently limits the retrievable documents to those with lexical matches. Though we argue that this is a reasonable starting point, it is not ideal to exclusively use such signals for dense retrieval because the highest-scoring results from a dense retrieval model need not have lexical overlap. LADR addresses this problem by expanding the scope of the search to the neighboring documents in the semantic space.

\noindent\textbf{HNSW}~\cite{8594636}. Both HNSW and LADR make use of document similarity scores when finding documents to score. While LADR makes use of an efficient lexical model to identify potentially relevant documents, HNSW uses a hierarchical structure to narrow into the neighborhood of the query vector. We argue that when strong and efficient lexical models are available to accelerate the selection process (like in many text retrieval settings), they should be used. Doing so allows LADR to allocate the compute to scoring more documents, rather than narrowing in on the neighborhood.

\noindent\textbf{GAR}~\cite{macavaney:cikm2022-adaptive}. LADR is perhaps most similar to Graph-Based Adaptive Re-ranking (GAR). Both techniques make use of an initial ranking and document similarities to select documents for scoring. GAR uses an alternating strategy to explore either the initial results, or exploit the results of the top-scoring document, with a major focus on prioritizing which to score next. This is important in the setting of expensive re-rankers, such as cross encoders, because the overall cost of scoring each document is very high (e.g., it's often only practical to score fewer than 1000 documents). Meanwhile, the cost of scoring a document in a bi-encoder setting is relatively cheap (it is reasonable to score a few thousand documents per query). Therefore, it is less important to be very selective about which documents to score, and instead broadly search in areas that have the \textit{potential} to yield relevant documents. This motivates the strategy of iterative exploration of neighbors of LADR. Indeed, in Section~\ref{rq1} we show that even when the GAR technique is properly optimized for a dense retrieval setting, it underperforms LADR.

\noindent\textbf{Hybrid Retrieval}~\cite{10.1007/978-3-030-72113-8_10,journals/corr/abs-2010-01195,lin-etal-2021-batch}. Ensembles of lexical and dense retrieval systems, often referred to as ``hybrid'' retrieval systems, also bear some semblance to LADR. Unlike hybrid retrieval systems, LADR only makes use of lexical retrieval systems insofar as to select a set of candidate documents, rather than using the lexical ranking score in the final order of the system itself. Consequently, LADR could be used in conjunction with a hybrid ensemble.

\noindent\textbf{Guided Traversal}~\cite{10.1145/3477495.3531774}. This strategy computes document scores for an arbitrary scoring function (in their work, a learned sparse scorer) for each document encountered while processing a lexical query. This can be seen as a generalization of the re-ranking strategy that also considers \textit{some}\footnote{In addition to the top $n$ lexical results, Guided Traversal scores documents encountered while traversing the posting lists for finding the top $n$ lexical results.} documents with lexical matches but not a high lexical ranking score. In contrast, LADR (1) is not constrained to only lexical matches; (2) avoids scoring documents that are neither high-scoring lexically nor neighbors of strong candidates; and (3) can easily augment existing retrieval pipelines, since it does not involve modifying the query traversal strategy, it simply takes the final output.

\section{Experimental Setup}\label{s:exp}

Via experimentation, we answer the following research questions:

\begin{enumerate}
\item[RQ1] How does LADR compare to other approximation techniques in terms of effectiveness and efficiency?
\item[RQ2] Is LADR applicable to a variety of single-representation dense retrieval models?
\item[RQ3] What are the computational overheads of LADR?
\item[RQ4] How do the parameters introduced by LADR (e.g., number of neighbors $k$) affect the effectiveness and efficiency of LADR?
\item[RQ5] Is an exact nearest neighbor graph needed for LADR to be effective?
\item[RQ6] What are the trade-offs between proactive and adaptive LADR in terms of precision, recall and latency?
\end{enumerate}

We have released the code to reproduce the results of our experiments.\footnote{http://github.com/Georgetown-IR-Lab/ladr}

\pageenlarge{1}
\subsection{Datasets and Measures}
\noindent\textbf{TREC 2019 Deep Learning (Passage Subtask).}
This is the official evaluation query set used in the TREC 2019 Deep Learning shared task \cite{https://doi.org/10.48550/arxiv.2003.07820}. It consists of 43 manually-judged queries using four relevance grades (215 relevance assessments per query, on average).

\noindent\textbf{TREC 2020 Deep Learning (Passage Subtask).}
This is the official evaluation query set used in the TREC 2020 Deep Learning shared task \cite{https://doi.org/10.48550/arxiv.2102.07662}. It consists of 54 queries with manual judgments from NIST annotators (211 relevance assessments per query, on average).  

\noindent\textbf{Dev (small).} This is the official small version of the dev set and has 6980 queries which is about 6.9\% of the full MS-Marco passage dev set. Unlike TREC DL 19/20, this dataset only contains a few known relevant passages per query (1.1 on average).
\\

We evaluate using nDCG and Recall (with a minimum relevance score of 2) at 1000 for TREC DL 2019 and 2020. Similarity, we evaluate using Mean Reciprocal Rank at 10 and Recall at 1000 for Dev (small). To measure how faithful the approximations are to the true rankings, we report Rank Biased Overlap (RBO~\cite{Webber2010ASM}) with $p=0.99$ of each method to the results from an exhaustive search.  We also report the per-query mean retrieval latency in milliseconds. Since query encoding time (i.e., the time to compute the query vector) is constant over all approximate nearest neighbor methods, we only consider the retrieval time in our measurements (i.e., we exclude query encoding latency).\footnote{Indeed, we also found that query encoding time exhibited higher variance across dense retrieval models than the the approximate nearest neighbor method.} Given the larger query sample, we use the Dev (small) dataset for all latency measurements.

\input{fig33}

\subsection{Models and Parameters}

We apply LADR and the baselines over the following diverse sample of single-representation dense retrieval models: TAS-B\footnote{\texttt{sebastian-hofstaetter/distilbert-dot-tas\_b-b256-msmarco}}~\cite{10.1145/3404835.3462891}, RetroMAE\footnote{\texttt{Shitao/RetroMAE\_MSMARCO\_distill}, we found our results to be slightly lower than the results reported in the paper because we use the official MS MARCO passage data, excluding the extra titles used in the RetroMAE experiments.}~\cite{RetroMAE}, TCT-ColBERT-HNP\footnote{\texttt{castorini/tct\_colbert-v2-hnp-msmarco}}~\cite{lin-etal-2021-batch}, and ANCE\footnote{\texttt{sentence-transformers/msmarco-roberta-base-ance-firstp}}~\cite{Xiong2021ApproximateNN}. We use BM25 over a BlockMax-WAND~\cite{10.1145/2009916.2010048} index encoding with default parameters using PISA~\cite{Mallia2019PISAPI} as our initial seed documents ($R_0$).

\input{tab_main}

For Proactive LADR, we primarily explore with a fixed $k=128$ (neighboring documents), varying $n\in[10,1000]$ (seed set size) to trade off efficiency and effectiveness. For Adaptive LADR, we primarily explore with a fixed $n=200$ and $k=128$, varying $c\in[1,200]$ (exploration depth) to trade off efficiency and effectiveness. We further explore the effects of $n$, $k$, and $c$ in Section~\ref{rq34}. By default, we use exact neighbors for LADR by preforming an exhaustive $k$ nearest neighbor search over all documents; we further explore the effect of using approximate or alternative graphs in Section~\ref{sec:res-doc-prox}.

\subsection{Baselines and Implementation}

We compare LADR with the following baselines:

\textbf{HNSW}~\cite{8594636} is a nearest neighbor search based indexing techniques which enables efficient dense retrieval. We use $k=128$ neighbors to match the settings of LADR, and vary the expansion factor in the range $[1,4096]$ to trade off effectiveness and efficiency.

\textbf{IVF}~\cite{1238663,DBLP:conf/eccv/YuanGCLJ12} is an inverted file index which is based on reduction of search scope using clustering. We evaluated IVF models with 4096 and 65536 centroids. IVF model with 4096 centroids was relatively efficient and delivered better results when compared with the one having 65536 centroids. We use between $[1,128]$ probes to trade off effectiveness and efficiency.

\textbf{Re-ranking} pipeline was also evaluated using BM25 for first-stage retrieval, re-scored using dense model. We trade off efficiency and effectiveness by varying $n\in[10,20000]$ (documents re-ranked).

\textbf{GAR}~\cite{macavaney:cikm2022-adaptive} alternates between exploring the initial results and the neighbors of the top-scoring documents. We use the same $k=128$ as with LADR, and vary the scoring budget in the range $[64,4096]$.

\textbf{ScaNN}~\cite{avq_2020} 
is a quantization-based technique which works on anisotropic quantization loss function for inner product search. We set the number of leaves to 4096 and search between $[1,128]$ of them to trade off effectiveness and efficiency.

For re-ranking, GAR, and LADR, we use our own implementations, leveraging the efficient dot product implementation provided by NumPy~\cite{harris2020array}.
For HNSW and IVF we use the FAISS implementation~\cite{johnson2019billion}. For ScaNN, we use the authors' implementation.\footnote{\url{https://github.com/google-research/google-research/tree/master/scann}} All methods run a single query at a time on a single CPU with the document vectors pre-loaded into memory. 
Experiemnts were conducted using AMD Ryzen Threadripper 1920X (3.9GHz) processor.

\pageenlarge{1}
\section{Results and Analysis}\label{s:exp}
    
This section presents results and analysis of different experiments with reference to our hypothesis and research questions. Figure~\ref{fig:maingraphs} presents the results of LADR and the baseline approaches across their effectiveness/efficiency trade-off parameters for TAS-B.
We start by exploring the relative strengths of the baselines. In terms of the precision-oriented measures, nDCG and RR@10, re-ranking and GAR out-perform HNSW, IVF, and ScaNN at most operational points. In terms of Recall@1k, we observe a similar trend, but with HNSW out-performing re-ranking and GAR above around 6ms/query on Dev (small), though not on TREC DL. Across all three datasets, HNSW produces result sets that are more faithful to exhaustive searches, as evidenced by far higher RBO scores than other baselines. Finally, we note that through our testing, we were unable to tune HNSW at or below 4ms/query with $k=128$.

\pageenlarge{1}
Next, we see that both LADR variants perform favorably to the baselines across all three measures. At low-latency settings (e.g., at or below 8ms/query), Adaptive LADR establishes a new Pareto frontier in terms of nDCG, RR@10, Recall@1k and RBO. However, akin to HNSW, we were unable to tune it such that it performed at or below 4ms/query with $k=128$; in these cases, the Proactive LADR variant is preferable. At higher latency settings, Adaptive LADR performs on-par with the best baselines across all datasets and measures (re-ranking/GAR for nDCG, RR@10, and Recall@1k, and HNSW for RBO). LADR achieves the ``best of both worlds'': higher effectiveness than the baselines at low latency settings, and effectiveness on par with the top baseline at high latency settings.

\pageenlarge{1}
\subsection{RQ1: nDCG and Recall@1k}
\label{rq1}
To establish whether LADR is truly more effective at low-latency settings, we conduct experiments at two operational points: 4ms/query and 8ms/query. For each variant of LADR and the baselines, we select parameters that yield latencies as close as possible to these values. We then apply the approach across four different dense retrieval models: TAS-B, RetroMAE, TCT-ColBERT-HNP and ANCE, and present the results in Table~\ref{tab:main}. For additional context, we also provide the performance of an exhaustive search for each of the models, though we note that these values do not correspond to the 4ms and 8ms operational points -- instead, they take around 60ms/query when using a modern GPU.

\begin{figure}[t]
\centering
\includegraphics[width=\linewidth]{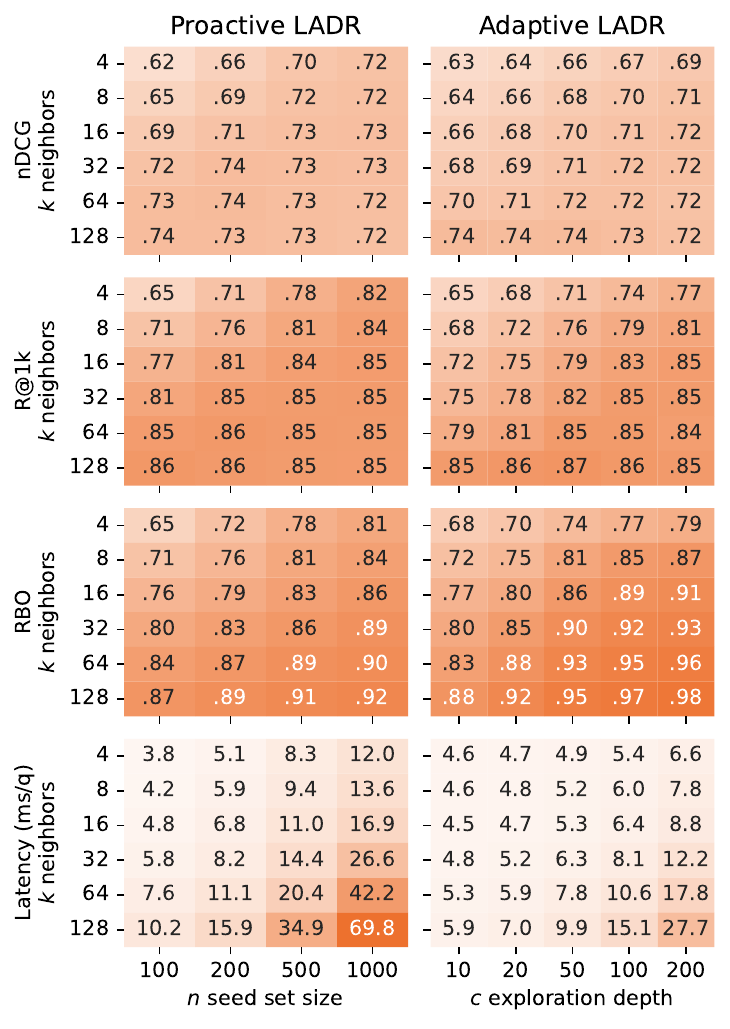}
\vspace{-3em}
\caption{Comparison of LADR performance over TAS-B on TREC DL 2019 while varying $k$, $n$, and $c$.}
\label{fig:rq3_rq4}
\end{figure}

When used with TAS-B, the baselines achieved at best an nDCG of 0.589 and 0.615 for DL 2019 and DL 2020 datasets with a 4ms time budget. These results were clearly outperformed by Proactive LADR with statistically significant improvement leading to a nDCG of 0.690 and 0.691, respectively.
Similarly with a 8ms budget for TAS-B, both Proactive and Adaptive LADR outperform the baselines with Adaptive LADR giving a nDCG of 0.738 (DL19) and 0.739 (DL20) over 0.688 and 0.699, the top baseline results.
The improvement in terms of Recall@1k is even more remarkable.
Recall@1k of 0.771 and 0.807 was observed over 0.605 and 0.667 (baselines) for Proactive LADR on DL 2019 and DL 2020 for a 4ms budget. Similarly, Recall@1k of 0.872 and 0.900 was observed over 0.755 and 0.836 (baselines) for Adaptive LADR on DL 2019 and DL 2020 for a 8ms budget.
On the Dev (small) dataset we observe that both Proactive and Adaptive LADR outperform the baselines in the Reciprocal Rank @10 and Recall@1k metrics in both 4ms and 8ms time budgets. We observe similar trends across other methods namely RetroMAE, TCT-ColBERT-HNP and ANCE. Another key observations is that both Proactive and Adaptive LADR with 8ms time budget yields higher nDCG and Recall@1k metrics than the exhaustive search across all dense retrieval methods on the DL 2019 and DL 2020 datasets. However, we note that this may be due, in part, to pooling bias, which could favor results included in the annotation pool (like BM25, which contributes to LADR).

These results answer RQ1: LADR consistently performs better than existing approximation techniques, both nDCG and Recall@1k, in low-latency settings. Further, it performs competitively with exhaustive retrieval, particularly at the 8ms/query operational point.

\pageenlarge{1}
\subsection{RQ2: Model Applicability} 

Table~\ref{tab:main} presents the Proactive and Adaptive LADR results across a variety of dense retrieval models. It shows statistically significant improvements over the baselines for DL19 and DL20 across all four dense retrieval methods, with the only exception being ANCE on DL20. These improvements are observed in both nDCG and Recall@1k metrics at multiple operational points. Further, this trend also largely holds on the Dev (small) dataset, though occasionally re-ranking and GAR achieve higher RR@10 performance (RetroMAE and ANCE). Hence, addressing RQ2 we conclude that both Proactive and Adaptive LADR are applicable to a variety of single-representation dense retrieval models.

\subsection{RQ3 and RQ4: Overhead and Parameters} \label{rq34}

Next, we perform an analysis of the parameters introduced by LADR: the number of neighbors per document in the proximity graph ($k$), the seed set size ($n$) and the exploration depth ($c$). Figure~\ref{fig:rq3_rq4} reports the nDCG, Recall@1k, RBO, and mean latency per query, when using the TAS-B model over the DL19 dataset.

We observe several trends, which should help researchers and practitioners tune LADR for their particular needs. First, as the number of neighbors in the proximity graph ($k$) increases, the performance in nDCG, Recall@1k and RBO (Rank Biased Overlap) also increase, but so does latency. This is expected, given that increasing $k$ results in LADR scoring more documents that are nearby potentially relevant ones~\cite{Jardine1971TheUO}. However, increasing $k$ comes at the cost of additional storage requirements; at $k=128$, 4.3GB of additional storage is required for the MS MARCO dataset. Gains can also be made when increasing the seed set size ($n$) or the exploration depth ($c$), though it is costlier in terms of compute (e.g., 12.0ms/q vs 10.2ms/q) for less gains (0.82 vs 0.86 Recall@1k). In terms of nDCG and Recall@1k, the gains from these parameters are complementary up to a point, which allows the user to balance their effectiveness, query-time efficiency, and storage requirements. Meanwhile, if the goal is to best approximate an exhaustive search, Adaptive LADR is the way to go; 0.98 RBO can be achieved at only 27.7ms/q when both a large $k$ and $c$ are used, compared to only 0.92 RBO for Proactive LADR, costing 69.8ms/q.
Thus, RQ3 and RQ4 have been addressed by studying changes in latency, effectiveness and efficiency with respect to parameters introduced by LADR, and adjusting these parameters allows the user to trade off effectiveness, query-time efficiency, and storage requirements.

\pageenlarge{1}
\subsection{RQ5: Document Proximity Sources}\label{sec:res-doc-prox}

Up to this point, we have explored the setting where the exact top $k$ nearest documents to each document in the target vector space are known. This is not always a realistic requirement, however; computing such scores is a quadratic operation over the number of documents in the collection. We therefore now explore two alternative sources of document proximity to test whether LADR can be used without requiring a precise similarity graph.

We explore two alternatives. First, we identify that HNSW~\cite{8594636} constructs its proximity graph in loglinear time by making use of approximate searches through its hierarchical structure for each new document. We therefore explore an \textit{Approximate} setting, wherein the document similarity graph is constructed by the HNSW index construction algorithm (Algorithm 1 in~\cite{8594636}). Meanwhile, \citet{macavaney:cikm2022-adaptive} found that using document proximity via BM25 provides similar effectiveness as dense vector document similarity in their adaptive re-ranking algorithm. Given that a BM25 graph would be model-agnostic, we also consider this strategy here.

We construct the two alternative document similarity graphs (Approx. and BM25), each linking up to $k=128$ nearby documents, and report the results in Table~\ref{tab:alt-sim} for TAS-B.
Table~\ref{tab:alt-sim} shows results for Exact proximity graph with TAS-B, Approx method (HNSW) and BM25 for the DL 2019, DL 2020 and Dev (small) datasets. As evident in Table~\ref{tab:alt-sim}, differences in nDCG and Recall@1k of Approx proximity graph when compared with Exact proximity graph are not significant for datasets DL 2019 and DL 2020. Similar trends are observed when BM25 based proximity graph and Exact proximity graph are compared. Further statistical equivalence for results without significant differences are also depicted in Table~\ref{tab:alt-sim}. Both Proactive and Adaptive LADR show high performance irrespective of the the proximity graph construction method.
The results suggest that both Proactive and Adaptive LADR are remarkably robust to these alternative proximity signals addressing RQ5.

\begin{table}
\centering
\small
\caption{Effectiveness of LADR over TAS-B when using alternative sources of document similarity, namely an approximation of the exact similarity graph from HNSW and a graph constructed using BM25. Operational points for Exact LADR are selected from the $\sim$8ms setting from Table~\ref{tab:main}; alternative sources use the same settings. Significant differences as compared to Exact are indicated with $*$ (paired t-test, $p<0.05$). For results without significant differences, we indicate statistical equivalence with $=$ (paired TOST, $p<0.05$, $|\Delta|<0.02$).}
\vspace{-1em}
\label{tab:alt-sim}
\begin{tabular}{lrrrrrr}
\toprule
& \multicolumn{2}{c}{DL19} &\multicolumn{2}{c}{DL20} & \multicolumn{2}{c}{Dev (sm)} \\
\cmidrule(lr){2-3}\cmidrule(lr){4-5}\cmidrule(lr){6-7}
Graph & nDCG & R@1k & nDCG & R@1k & RR@10 & R@1k \\

\midrule
\multicolumn{7}{c}{\bf Proactive LADR} \\
\midrule
Exact & 0.730 &\bf  0.850 &\bf  0.722 &\bf  0.857 &\bf  0.345 &\bf  0.932 \\
Approx. & $^{=}_{}$0.731 & 0.845 & $^{=}_{}$0.720 & 0.849 & $^{*}_{}$0.343 & $^{*}_{}$0.916 \\
BM25 &\bf  $^{=}_{}$0.732 & 0.835 & $^{=}_{}$0.720 & 0.853 & $^{*}_{}$0.339 & $^{*}_{}$0.883 \\
\midrule
\multicolumn{7}{c}{\bf Adaptive LADR} \\
\midrule
Exact & 0.738 &\bf  0.872 & 0.739 &\bf  0.900 &\bf  0.347 & 0.960 \\
Approx. & $^{=}_{}$0.736 & 0.861 & $^{=}_{}$0.737 &\bf  $^{=}_{}$0.900 &\bf  $^{=}_{}$0.347 &\bf  $^{*}_{}$0.966 \\
BM25 &\bf  0.743 & 0.859 &\bf  $^{=}_{}$0.742 &\bf  0.900 & $^{*}_{}$0.345 & $^{*}_{}$0.933 \\

\bottomrule
\end{tabular}
\end{table}

\subsection{RQ6: Proactive and Adaptive Trade-offs}

Adaptive LADR performs better than Proactive LADR in the metrics of nDCG and Recall@1k in the same time budget. Adaptive LADR by being more selective about neighbor selection through an iterative approach ensures that the limited computational resources are being spent on promising documents. One limitation of Adaptive LADR is that operational points with very low time budget do not exist. Additionally, Proactive LADR has O(kn) worst case time complexity where k is the number of neighbors and n is the number of seed documents. While for Adaptive LADR worst case time complexity is O(D) where D is the number of documents in the collection. Even though worst case is unlikely, latency for Adaptive LADR will be at par with Exhaustive retrieval. Considering the trade-offs between Proactive and Adaptive LADR for very low operational points Proactive LADR would be a better choice. While, Adaptive LADR with a timeout strategy would be a better choice for cases with relatively higher time budgets addressing RQ6.

\pageenlarge{1}  
\section{Conclusions and Future Directions}\label{s:conc}
The ability to identify related documents that do not necessarily contain same terms but are closer in the semantic space is the key of dense retrieval approaches. 
This requires exhaustive search which is an expensive operation.
Current state of the art approximate k nearest neighbor methods like HNSW, ScaNN, IVF etc try to provide some resolve to this challenging task but fail to match the precision and recall of exhaustive search.
Our simple-yet-effective method `LADR’ (Lexically-Accelerated Dense Retrieval) uses lexical retrieval techniques to seed a dense retrieval exploration that uses a document proximity graph. 
Both Proactive and Adaptive LADR establish a new dense retrieval effectiveness-efficiency Pareto frontier among approximate k nearest neighbor techniques.
While Proactive LADR outperforms baselines at lower time budgets, Adaptive LADR focuses on optimal use of compute on promising documents leading to a major leap in recall.
LADR consistently achieves both nDCG and recall that are on par with an exhaustive search on standard benchmarks without needing GPUs.
 
One natural area for further exploration is in developing new LADR variants; Proactive and Adaptive are appealing in their simplicity and remarkable effectiveness, but there likely exist smarter exploration strategies. Given that LADR relies on intra-document similarity, another interesting direction would be to train dense retrieval models with this quality in mind -- e.g., by introducing a new intra-document similarity objective in the training process, rather than query-document similarity alone.

\bibliographystyle{ACM-Reference-Format}
\bibliography{sample-base}

\end{document}

%% file: fig33.tex
\begin{figure*}
\centering
TREC DL 2019

\includegraphics[scale=0.55]{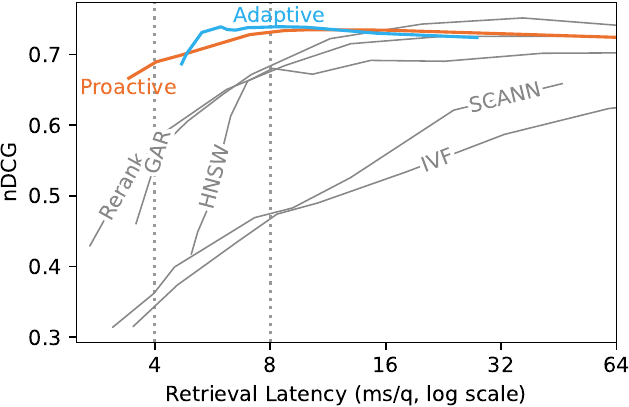}
\includegraphics[scale=0.55]{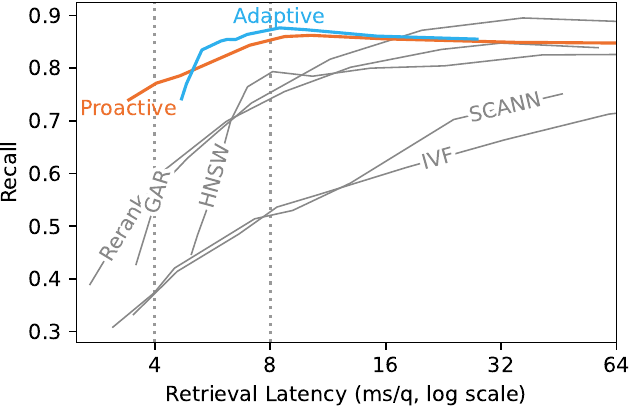}
\includegraphics[scale=0.55]{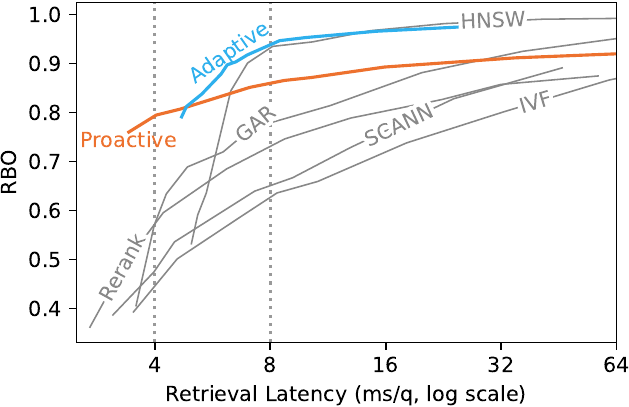}

TREC DL 2020

\includegraphics[scale=0.55]{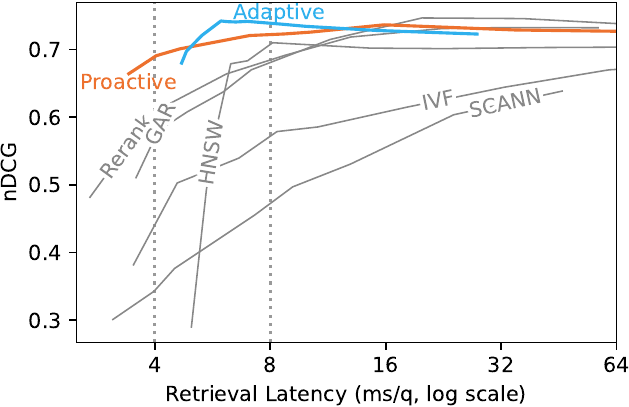}
\includegraphics[scale=0.55]{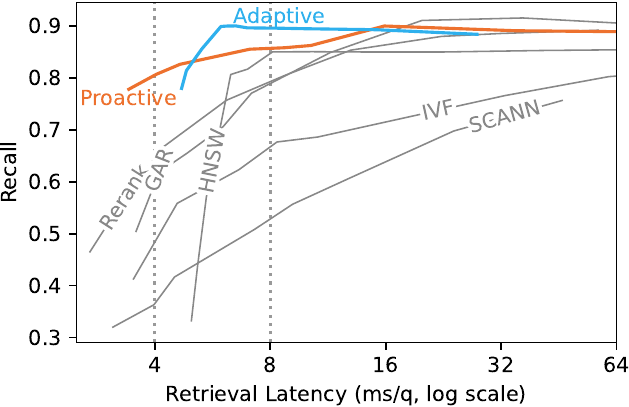}
\includegraphics[scale=0.55]{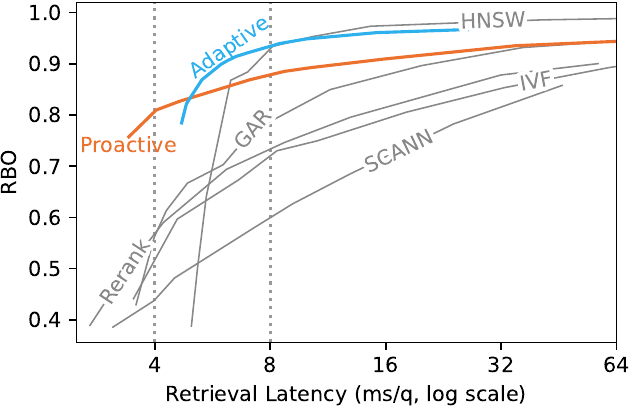}

MS MARCO Dev (small)

\includegraphics[scale=0.55]{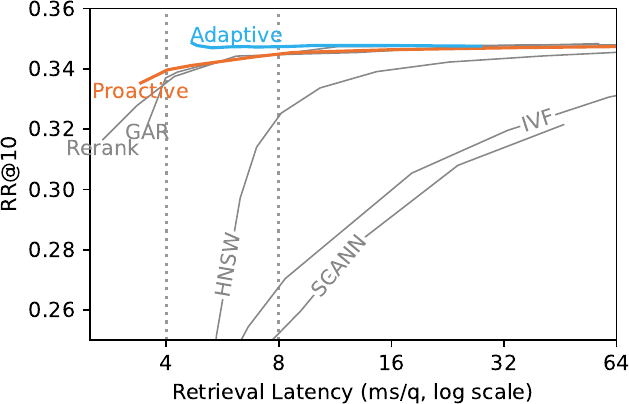}
\includegraphics[scale=0.55]{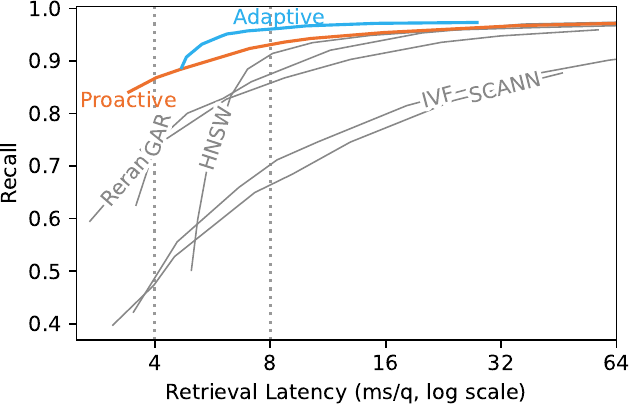}
\includegraphics[scale=0.55]{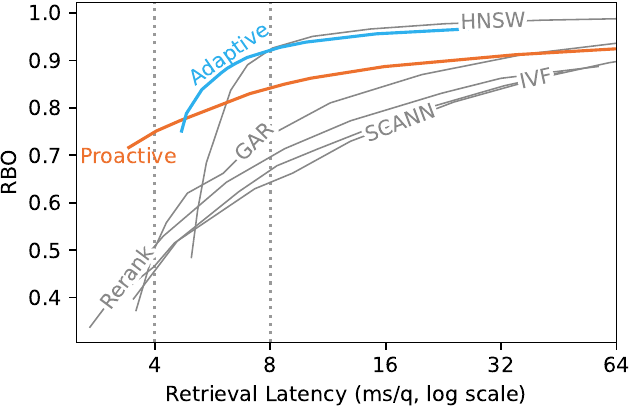}
\vspace{-1em}
\caption{Performance of LADR over TAS-B and baselines across various operational points.}
\label{fig:maingraphs}
\end{figure*}

%% file: tab_main.tex
\begin{table*}
\centering
\renewcommand{\arraystretch}{1.1}
\small
\caption{Retrieval effectiveness comparison of various dense scoring models and approximate retrieval methods at two operational points (approximately 4ms/query and 8ms/query) for TREC DL19, TREC DL20, and MS MARCO Dev (small). Results that cannot be matched to the operational point are indicated with ``-''. Significant improvements between approaches and LADR are indicated using superscript/subscript indicators (paired t-test, $p<0.05$). The indicators are $I$, $S$, $H$, $G$ and $R$ that correspond to respective baselines. For context, exhaustive retrieval effectiveness is also provided for each model in \textcolor{gray}{gray}, though note that these values do not correspond with the operational points.}
\label{tab:main}
\scalebox{0.92}{
\begin{tabular}{l|rrrr|rrrr|rrrr}
\toprule
& \multicolumn{2}{c}{DL19 $\sim$4ms} & \multicolumn{2}{c|}{DL19 $\sim$8ms} & \multicolumn{2}{c}{DL20 $\sim$4ms} & \multicolumn{2}{c|}{DL20 $\sim$8ms} & \multicolumn{2}{c}{Dev (sm) $\sim$4ms} & \multicolumn{2}{c}{Dev (sm) $\sim$8ms} \\
\cmidrule(lr){2-3}
\cmidrule(lr){4-5}
\cmidrule(lr){6-7}
\cmidrule(lr){8-9}
\cmidrule(lr){10-11}
\cmidrule(lr){12-13}
Method & nDCG & R@1k & nDCG & R@1k & nDCG & R@1k & nDCG & R@1k & RR@10 & R@1k & RR@10 & R@1k \\

\midrule
\midrule
\textbf{TAS-B} (Exh.) & \textcolor{gray}{0.715} & \textcolor{gray}{0.842} & \textcolor{gray}{0.715} & \textcolor{gray}{0.842} & \textcolor{gray}{0.713} & \textcolor{gray}{0.875} & \textcolor{gray}{0.713} & \textcolor{gray}{0.875} & \textcolor{gray}{0.347} & \textcolor{gray}{0.978} & \textcolor{gray}{0.347} & \textcolor{gray}{0.978} \\
\midrule
IVF [$I$] & 0.374 & 0.414 & 0.474 & 0.536 & 0.503 & 0.559 & 0.579 & 0.677 & 0.217 & 0.556 & 0.270 & 0.712 \\
ScaNN [$S$] & 0.475 & 0.519 & 0.537 & 0.598 & 0.476 & 0.527 & 0.553 & 0.641 & 0.254 & 0.669 & 0.292 & 0.774 \\
HNSW [$H$] & - & - & 0.614 & 0.707 & - & - & 0.699 & 0.836 & - & - & 0.310 & 0.872 \\
GAR [$G$] & 0.543 & 0.540 & 0.688 & 0.755 & 0.568 & 0.594 & 0.684 & 0.796 & 0.337 & 0.732 & 0.345 & 0.876 \\
Re-Ranking [$R$] & 0.589 & 0.605 & 0.684 & 0.755 & 0.615 & 0.667 & 0.691 & 0.805 & 0.337 & 0.748 & 0.345 & 0.868 \\
Proactive LADR &\bf  $^{IS}_{GR}$0.690 &\bf  $^{IS}_{GR}$0.771 & $^{ISH}_{GR}$0.730 & $^{ISH}_{GR}$0.850 &\bf  $^{IS}_{GR}$0.691 &\bf  $^{IS}_{GR}$0.807 & $^{IS}_{GR}$0.722 & $^{IS}_{GR}$0.857 &\bf  $^{IS}_{}$0.340 &\bf  $^{IS}_{GR}$0.868 & $^{ISH}_{}$0.345 & $^{ISH}_{GR}$0.932 \\
Adaptive LADR & - & - &\bf  $^{ISH}_{GR}$0.738 &\bf  $^{ISH}_{GR}$0.872 & - & - &\bf  $^{ISH}_{GR}$0.739 &\bf  $^{ISH}_{GR}$0.900 & - & - &\bf  $^{ISH}_{GR}$0.347 &\bf  $^{ISH}_{GR}$0.960 \\
\midrule
\midrule
\textbf{RetroMAE} (Exh.) & \textcolor{gray}{0.699} & \textcolor{gray}{0.806} & \textcolor{gray}{0.699} & \textcolor{gray}{0.806} & \textcolor{gray}{0.701} & \textcolor{gray}{0.839} & \textcolor{gray}{0.701} & \textcolor{gray}{0.839} & \textcolor{gray}{0.375} & \textcolor{gray}{0.981} & \textcolor{gray}{0.375} & \textcolor{gray}{0.981} \\
\midrule
IVF [$I$] & 0.226 & 0.225 & 0.346 & 0.358 & 0.272 & 0.263 & 0.372 & 0.375 & 0.157 & 0.381 & 0.221 & 0.541 \\
ScaNN [$S$] & 0.468 & 0.502 & 0.525 & 0.588 & 0.486 & 0.509 & 0.555 & 0.606 & 0.275 & 0.665 & 0.312 & 0.769 \\
HNSW [$H$] & - & - & 0.630 & 0.720 & - & - & 0.673 & 0.798 & - & - & 0.338 & 0.874 \\
GAR [$G$] & 0.559 & 0.553 & 0.696 & 0.763 & 0.578 & 0.604 & 0.692 & 0.789 &\bf  0.357 & 0.750 & 0.368 & 0.890 \\
Re-Ranking [$R$] & 0.594 & 0.605 & 0.685 & 0.755 & 0.622 & 0.667 & 0.696 & 0.805 & 0.355 & 0.748 & 0.369 & 0.868 \\
Proactive LADR &\bf  $^{IS}_{GR}$0.691 &\bf  $^{IS}_{GR}$0.765 & $^{ISH}_{GR}$0.733 & $^{ISH}_{GR}$0.844 &\bf  $^{IS}_{GR}$0.702 &\bf  $^{IS}_{GR}$0.811 & $^{ISH}_{G}$0.723 & $^{IS}_{G}$0.846 & $^{IS}_{}$0.356 &\bf  $^{IS}_{GR}$0.864 & $^{ISH}_{}$0.368 & $^{ISH}_{GR}$0.938 \\
Adaptive LADR & - & - &\bf  $^{ISH}_{R}$0.740 &\bf  $^{ISH}_{GR}$0.866 & - & - &\bf  $^{ISH}_{G}$0.731 &\bf  $^{ISH}_{GR}$0.879 & - & - &\bf  $^{ISH}_{GR}$0.374 &\bf  $^{ISH}_{GR}$0.973 \\
\midrule
\midrule
\textbf{TCT-HNP} (Exh.) & \textcolor{gray}{0.708} & \textcolor{gray}{0.830} & \textcolor{gray}{0.708} & \textcolor{gray}{0.830} & \textcolor{gray}{0.689} & \textcolor{gray}{0.848} & \textcolor{gray}{0.689} & \textcolor{gray}{0.848} & \textcolor{gray}{0.359} & \textcolor{gray}{0.970} & \textcolor{gray}{0.359} & \textcolor{gray}{0.970} \\
\midrule
IVF [$I$] & 0.340 & 0.366 & 0.437 & 0.469 & 0.369 & 0.383 & 0.470 & 0.522 & 0.219 & 0.527 & 0.276 & 0.687 \\
ScaNN [$S$] & 0.378 & 0.410 & 0.444 & 0.496 & 0.355 & 0.376 & 0.427 & 0.459 & 0.215 & 0.522 & 0.253 & 0.632 \\
HNSW [$H$] & - & - & 0.625 & 0.721 & - & - & 0.634 & 0.762 & - & - & 0.315 & 0.853 \\
GAR [$G$] & 0.546 & 0.547 & 0.687 & 0.755 & 0.569 & 0.598 & 0.678 & 0.797 & 0.342 & 0.733 & 0.354 & 0.878 \\
Re-Ranking [$R$] & 0.586 & 0.605 & 0.679 & 0.755 & 0.614 & 0.667 & 0.685 & 0.805 & 0.342 & 0.748 & 0.353 & 0.868 \\
Proactive LADR &\bf  $^{IS}_{GR}$0.680 &\bf  $^{IS}_{GR}$0.747 & $^{ISH}_{GR}$0.719 & $^{ISH}_{GR}$0.827 &\bf  $^{IS}_{GR}$0.682 &\bf  $^{IS}_{GR}$0.803 & $^{ISH}_{G}$0.709 & $^{ISH}_{}$0.841 &\bf  $^{IS}_{G}$0.346 &\bf  $^{IS}_{GR}$0.856 & $^{ISH}_{}$0.354 & $^{ISH}_{GR}$0.927 \\
Adaptive LADR & - & - &\bf  $^{ISH}_{}$0.729 &\bf  $^{ISH}_{GR}$0.848 & - & - &\bf  $^{ISH}_{GR}$0.721 &\bf  $^{ISH}_{GR}$0.878 & - & - &\bf  $^{ISH}_{GR}$0.359 &\bf  $^{ISH}_{GR}$0.962 \\
\midrule
\midrule
\textbf{ANCE} (Exh.) & \textcolor{gray}{0.617} & \textcolor{gray}{0.755} & \textcolor{gray}{0.617} & \textcolor{gray}{0.755} & \textcolor{gray}{0.634} & \textcolor{gray}{0.777} & \textcolor{gray}{0.634} & \textcolor{gray}{0.777} & \textcolor{gray}{0.330} & \textcolor{gray}{0.957} & \textcolor{gray}{0.330} & \textcolor{gray}{0.957} \\
\midrule
IVF [$I$] & 0.358 & 0.395 & 0.441 & 0.500 & 0.407 & 0.437 & 0.498 & 0.549 & 0.212 & 0.530 & 0.268 & 0.703 \\
ScaNN [$S$] & 0.374 & 0.405 & 0.433 & 0.488 & 0.440 & 0.495 & 0.535 & 0.614 & 0.262 & 0.691 & 0.287 & 0.783 \\
HNSW [$H$] & - & - & 0.606 & 0.737 & - & - & 0.635 & 0.790 & - & - & 0.311 & 0.897 \\
GAR [$G$] & 0.527 & 0.540 & 0.648 & 0.750 & 0.568 & 0.622 & 0.655 & 0.794 &\bf  0.326 & 0.751 & 0.329 & 0.888 \\
Re-Ranking [$R$] & 0.578 & 0.605 & 0.653 & 0.755 & 0.602 & 0.667 &\bf  0.674 & 0.805 & 0.325 & 0.748 &\bf  0.333 & 0.868 \\
Proactive LADR &\bf  $^{IS}_{GR}$0.645 &\bf  $^{IS}_{GR}$0.751 & $^{IS}_{}$0.657 & $^{ISH}_{}$0.800 &\bf  $^{IS}_{GR}$0.660 &\bf  $^{IS}_{GR}$0.807 & $^{IS}_{}$0.666 & $^{IS}_{}$0.822 & $^{IS}_{}$0.321 &\bf  $^{IS}_{GR}$0.872 & $^{ISH}_{}$0.327 & $^{ISH}_{GR}$0.932 \\
Adaptive LADR & - & - &\bf  $^{ISH}_{}$0.665 &\bf  $^{ISH}_{}$0.820 & - & - & $^{ISH}_{}$0.665 &\bf  $^{IS}_{}$0.830 & - & - & $^{ISH}_{}$0.329 &\bf  $^{ISH}_{GR}$0.959 \\

\bottomrule
\end{tabular}}
\end{table*}